\documentclass[sigconf]{acmart}

\usepackage[utf8]{inputenc}
\usepackage[english]{babel}
\usepackage{graphicx}
\usepackage{listings}
\usepackage{array}
\usepackage{booktabs}
\usepackage{multirow}
\usepackage{siunitx}
\usepackage{multirow}
\usepackage{amsmath}
\usepackage{pdfpages}
\usepackage{float}
\usepackage{booktabs}
\usepackage{dblfloatfix}
\usepackage{dsfont}

\begin{document}
\title{Recommendation of Move Method Refactoring Using Path-Based Representation of Code}

\author{Zarina Kurbatova}
\affiliation{JetBrains Research\\
Saint-Petersburg State University, Russia}
\email {zarina.kurbatova@gmail.com}

\author{Ivan Veselov}
\affiliation{Higher School of Economics, Russia}
\email {veselov.iva@gmail.com}

\author{Yaroslav Golubev}
\affiliation{JetBrains Research\\
ITMO University, Russia}
\email{golubev@itmo.ru}

\author{Timofey Bryksin}
\affiliation{JetBrains Research\\
Saint-Petersburg State University, Russia}
\email{t.bryksin@spbu.ru}

\begin{abstract}
Software refactoring plays an important role in increasing code quality. One of the most popular refactoring types is the Move Method refactoring. It is usually applied when a method depends more on members of other classes than on its own original class. Several approaches have been proposed to recommend Move Method refactoring automatically. Most of them are based on heuristics and have certain limitations (e.g., they depend on the selection of metrics and manually-defined thresholds). In this paper, we propose an approach to recommend Move Method refactoring based on a path-based representation of code called \textit{code2vec} that is able to capture the syntactic structure and semantic information of a code fragment. We use this code representation to train a machine learning classifier suggesting to move methods to more appropriate classes. We evaluate the approach on two publicly available datasets: a manually compiled dataset of well-known open-source projects and a synthetic dataset with automatically injected code smell instances. The results show that our approach is capable of recommending accurate refactoring opportunities and outperforms JDeodorant and JMove, which are state of the art tools in this field.
\end{abstract}

\maketitle
\section{Introduction}
Beck coined a metaphor of \textit{code smell}~\cite{fowler2018refactoring} that describes possible cases of poor design or inappropriate implementation choices. Smelly code is usually harder to read and understand. Fowler and Beck~\cite{beck1999bad} introduced twenty two smells as indicators of some deeper problems in code, one of the most common of them is \textit{Feature Envy}. This smell relates to a situation when a method is more interested in the content or behaviour of another class than in its original class. The authors also provided a refactoring strategy for each smell --- a process that improves the internal structure of software applications while leaving their behavior unchanged~\cite{fowler2018refactoring}. Refactoring tends to make software easier to understand and maintain. For instance, the \textit{Move Method} refactoring is considered to be a solution to the Feature Envy smell: moving the method to the class to which it is more closely related. The main benefit of applying Move Method refactoring is the reduction in coupling between classes, which usually makes code more flexible.

Manually checking the source code to identify refactoring opportunities is a tiresome and time-consuming process. Over the last years, multiple tools for automatic recommendation of refactoring opportunities have been introduced~\cite{tsantalis2009identification,terra2018jmove,palomba2013detecting,palomba2016textual, moha2009decor,ghannem2016use,liu2016domino,ujihara2017c, saranya2018model,bryksin2018automatic,palomba2014mining}, some of them are implemented as plugins for integrated development environments (IDEs), while others are stand-alone tools. A detailed review and comparison of existing approaches has been recently compiled by Pecorelli et al.~\cite{pecorelli2019comparing}. Most of the existing Move Method refactoring recommendation approaches are based on manually-defined heuristics, which come to life when researchers try to formalize the meaning of high-quality code. Heuristic-based approaches are rather simple to define and implement: they calculate a set of software metrics and then use certain thresholds to determine whether the code is smelly or not. This leads to significant limitations, such as threshold dependability and subjective interpretation by developers, as well as a very low agreement between these approaches~\cite{fontana2012automatic}. To overcome these limitations, machine learning techniques have been employed. Fontana et al.~\cite{fontana2016comparing} showed that using classification algorithms is a promising way of detecting code smells. Nevertheless, some machine learning-based approaches continue to use metrics that carry the same limitations with them.

In this paper, we propose an approach to suggest Move Method refactoring opportunities based on measuring semantic similarity between a method and a class. To capture semantics of the code, we rely on its path-based representation that can be mapped to continuous distributed vectors, called embeddings. Such code embeddings are built in a way that allows to map semantically similar code snippets to close vectors. This technique has already been successfully used in the task of method name prediction~\cite{alon2019code2vec}, which makes it a valid candidate for the refactoring recommendation task as well.

The approach consists in searching for a potentially movable method, gathering a set of classes that this method can be moved to (including the possibility of it staying in its original class) and creating their path-based representation. After that, we employ a trained classifier that recommends to move a method to a more semantically similar class or leave it in its place if the most appropriate class is the current one. 
 
We also report the results of the evaluation, divided into two parts: (1) evaluation on a manually compiled dataset introduced in~\cite{terra2013qualitas}, and (2) evaluation on a synthetic dataset of projects with automatically injected Feature Envy smell instances that was introduced in~\cite{Novozhilov:2019:EMM:3340633.3340640}. In both cases, the comparison shows that our approach outperforms state-of-the-art tools JDeodorant~\cite{tsantalis2009identification} and JMove\cite{terra2018jmove}.

Therefore, our contributions are:
\begin{itemize}
    \item We introduce the idea of using path-based representation of code to build machine learning models for the task of refactoring recommendation.
    \item We implement the proposed approach to identify Feature Envy code smells and recommend appropriate Move Method refactoring using a machine learning technique called Support Vector Machine (SVM).
    \item We evaluate our approach on two publicly available datasets. For the first dataset the approach demonstrates an increase of 32\% in $F_1$ score compared to JMove, and an increase of 57\% compared to JDeodorant. For the second dataset it achieves a two-fold increase in $F_1$ score compared to JMove.
\end{itemize}

The paper is structured as follows. Section~\ref{section:relatedwork} provides a brief survey of refactoring recommendation approaches, Section~\ref{section:approach} describes the proposed approach, Section~\ref{section:evaluation} presents evaluation results. Section~\ref{section:threats} discusses possible threats to the validity of our research and Section~\ref{section:conclusion} makes conclusions and provides several directions for future work.

\begin{figure*}
  \centering
  \includegraphics[width=1\textwidth]{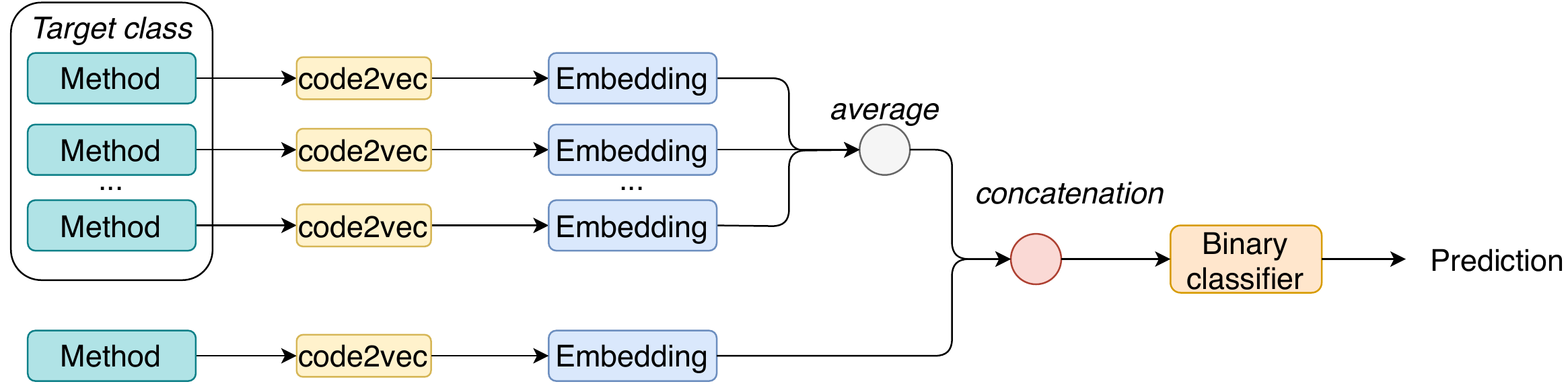}
  \caption{Overview of the proposed approach}
  \label{fig:proposed_approach}
\end{figure*}

\section{Related work}
\label{section:relatedwork}
Over the last decades, a number of approaches have been proposed to identify Feature Envy and recommend Move Method refactoring automatically. In this paper, we focus on comparing heuristics based approaches and algorithms that employ machine learning techniques to recommend refactorings. In this section, we provide an overview of existing approaches and broadly classify them into three categories: the first ones are based on various heuristics and heavily rely on software metrics, the second employ machine learning algorithms, and the third ones are merging results of other approaches using ensemble methods.

\subsection{Heuristics based approaches}
\label{section:heuristics-based}
The most well-known tool in this area is JDeodorant, which was introduced by Tsantalis and Chatzigeorgiou~\cite{tsantalis2009identification}. The tool is able to detect several code smells, namely, Feature Envy, Long Method, Type/State Checking, and God Class, and make appropriate refactoring suggestions to resolve them. Following its definition, the detection of Feature Envy is based on a heuristic that a method should be moved to a class where it accesses more entities (fields and methods) than it does in its own class. To make suggestions more reliable, the authors introduce a set of preconditions that should be satisfied for a Move Method refactoring to be recommended, for example, there are compilation preconditions that ensure that the code will be compiled correctly after the modification.

Terra et al.~\cite{terra2018jmove} presented a tool called JMove for the recommendation of Move Method refactoring based on similarity between dependency sets. The authors consider dependencies such as method calls, field accesses, return types, object instantiations, etc. The approach is based on an assumption that methods in well-designed classes usually establish dependencies to similar types. The dependencies established by a method are compared with the dependencies established by the methods in a possible target class. The authors evaluated the approach on 10 open-source projects. The evaluation is carried out separately for small and large methods, because JMove does not provide recommendations for methods with less than four dependencies. The results show that JMove outperforms JDeodorant and tends to work better on large methods. The authors also measured the performance of the tools, and results show that on average JMove takes more than three hours to perform the necessary computations and recommend refactorings, whereas JDeodorant takes 48 minutes, indicating that JMove takes about four times longer. 

Liu et al.~\cite{liu2016domino} proposed an approach based on other method movements. The intuition behind the proposed method is that similar methods should be moved together. Whenever a method is moved, the tool checks methods within the same class and suggests to move methods that have a strong relationship with the moved method. The strength of the relationship between methods is measured by computing several metrics such as coupling, conceptual correlation, and their similarity in distance which is calculated using distance metrics introduced in~\cite{tsantalis2009identification}.

Ujihara et al.~\cite{ujihara2017c} presented a refactoring recommendation tool based on semantic and static program analysis. The authors capture semantic similarity between a method and a class by computing cosine similarity between TF-IDF vectors. The authors assume that a method $m$ should be moved from a class $C_s$ to a class $C_t$ if $m$ is more similar to methods in $C_t$ than to methods in $C_s$. To identify refactoring opportunities, the approach generates a dependency graph for the source code and calculates a metric that takes into account such features as the number of edges/client classes/dependent classes to be added or removed when applying Move Method refactoring.

Bavota et al.~\cite{bavota2013methodbook} proposed an approach to identify Move Method opportunities using Relational Topic Models (RTM). RTM is a model of documents and links between them. The approach considers both structural (method calls) and textual (identifiers and words in comments) information extracted from the source code. 

Palomba et al.~\cite{palomba2016textual} presented an approach to detect code smells called TACO (Textual Analysis for Code Smell Detection). The approach is based on information retrieval methods and is able to detect Feature Envy, Long Method, Blob, Promiscuous Package, and Misplaced Class code smells. 

\subsection{Machine learning-based approaches}
\label{section:ml-based}
Liu et al.~\cite{8807230} proposed a deep learning based approach to detect Feature Envy, Long Method, Large Class, and Misplaced Class code smells. To identify Feature Envy, the authors extract textual information (e.g., identifier names) from source code, then map it to continuous distributed vectors (identifier embeddings) using the $word2vec$ technique~\cite{mikolov2013efficient}. Also, the authors calculate the distance between a method and a target class, and the distance between a method and its original class. These values are fed into a Convolutional Neural Network (CNN). The authors employ a CNN because CNNs may learn semantic relationships among the identifiers that may be useful to determine where the method should be placed. Apart from that, the authors implement well-known metrics from others studies: for example, to identify Feature Envy, they use metrics proposed by Tsantalis and Chatzigeorgiou~\cite{tsantalis2009identification}. 

Hadj-Kacem and Bouassida~\cite{hadj2019deep} have also proposed an approach to detect code smells (Feature Envy, Blob, and Long Method) using deep learning techniques. The authors parse the source code into Abstract Syntax Trees and then convert each tree into a vector representation using a \textit{coding criterion} proposed in~\cite{mou2014building}. According to the coding criterion, the vector for each non-leaf node in AST is calculated using vector representations of its children. Then, they fed the resulting vectors to a Variational Auto-Encoder model. To determinate whether the method should be moved or not, the authors use a Linear Regression classifier. The approach has been evaluated on 20 open-source projects and compared with TACO~\cite{palomba2016textual}, which it managed to outperform.


Sharma et al.~\cite{sharma2019feasibility} also presented the results of using deep learning techniques to detect code smells (Complex Method, Empty Catch Block, Magic Number, and Multifaceted Abstraction). To gather training data, the authors use Designite~\cite{sharma2016designite}, the software design quality assessment tool that supports the detection of 19 design code smells. The results show that a Recurrent Neural Network (RNN) performs better than a CNN for detecting Empty Catch Block and Magic Number code smells. The authors also investigated the effectiveness of the transfer-learning technique by training a deep learning model on samples of C\# code and then using this model to predict code smells in Java. The results show that transfer-learning technique is feasible for code smells detection with performance comparable to that of direct-learning.

\subsection{Ensemble methods}
Ensemble methods combine results from several algorithms to provide better suggestions.

Bryksin et al.~\cite{bryksin2018automatic} proposed an approach to recommend Move Method refactorings using clustering ensembles. The approach combines the results of several heuristic-based algorithms, such as ARI (Automatic Refactoring Identification)~\cite{marian2012using}, HAC (Hierarchical Agglomerative Clustering)~\cite{marian2012study}, and CCDA (Constrained Community Detection Algorithm)~\cite{pan2013refactoring}. Based on this approach, the authors presented a tool called ArchitectureReloaded that is implemented as a plug-in for IntelliJ IDEA and allows to run the selected algorithms and automatically recommend Move Method refactoring.

In a recent study, Barbez et al.~\cite{barbez2020machine} have also proposed a machine learning based ensemble method that combines multiple tools to detect code smells such as Feature Envy and God Class. The core idea of their method is to obtain software metrics from several approaches for each input entity and use them to train a machine learning classifier. To detect Feature Envy, the authors selected seven metrics from the following code smell detectors: JDeodorant~\cite{tsantalis2009identification}, InCode~\cite{marinescu2010incode}, and HIST~\cite{palomba2013detecting}. The evaluation results show that the proposed approach outperforms stand-alone tools that were aggregated.

In general, ensemble-based methods show better results than stand-alone tools because they are trying to combine the best parts from other existing approaches. As for the approaches themselves, techniques based on machine learning techniques look more promising because they can avoid the above-mentioned limitations and showcase better results than approaches that are based on heuristics and crafting features by hand.

\section{Approach}
\label{section:approach}
In this section, we describe our approach for recommending Move Method refactoring in greater detail. Firstly, we provide an overview of the approach, describe a way to build vector representations for code fragments that captures their semantic properties, and then we describe an adopted machine learning technique and the generation of training data for it.

\subsection{Overview}
\label{subsection:overview}
To recommend a refactoring, we first compile a set of potentially movable methods $m = \{m_1, m_2, ..., m_n\}$ and corresponding target classes for each method $tc_i = \{c_{i_1}, c_{i_2}, ..., c_{i_k}\}$ where the method could be moved to. The original class of the method is also considered as a target class. After that, for each method $m_i$, we convert it and a corresponding target class $c_{i_j}$ into numerical vectors and concatenate those two vectors into a single vector $v_{i_j}$. We feed each of the resulting vectors $v_{i_j}$ to a probabilistic classifier and set a certain threshold to conclude whether a method $m_i$ should be moved to $c_{i_j}$ or not. After that, we recommend to move the method $m_{i}$ to a class with the highest probability. If a class with the highest probability is an enclosing class of the method, we conclude that the method $m_i$ should not be moved. We also do not recommend refactoring if none of the possible moves have a probability larger than the threshold in order not to suggest unreliable refactorings.

The structure of the proposed model is presented in Figure~\ref{fig:proposed_approach}.

\subsection{Path-based representation of code}
\label{code2vec}
To make use of machine learning techniques when working with source code, we first have to get a numerical representation of the code fragments. To obtain such a representation, Alon et al.~\cite{alon2019code2vec} proposed a neural technique called \textit{code2vec}, which learns distributed vectors (embeddings) of code fragmants using their syntactic structure and semantic information that is captured in names of methods and identifiers. 

\begin{figure}
  \centering
  \includegraphics[width=0.5\textwidth]{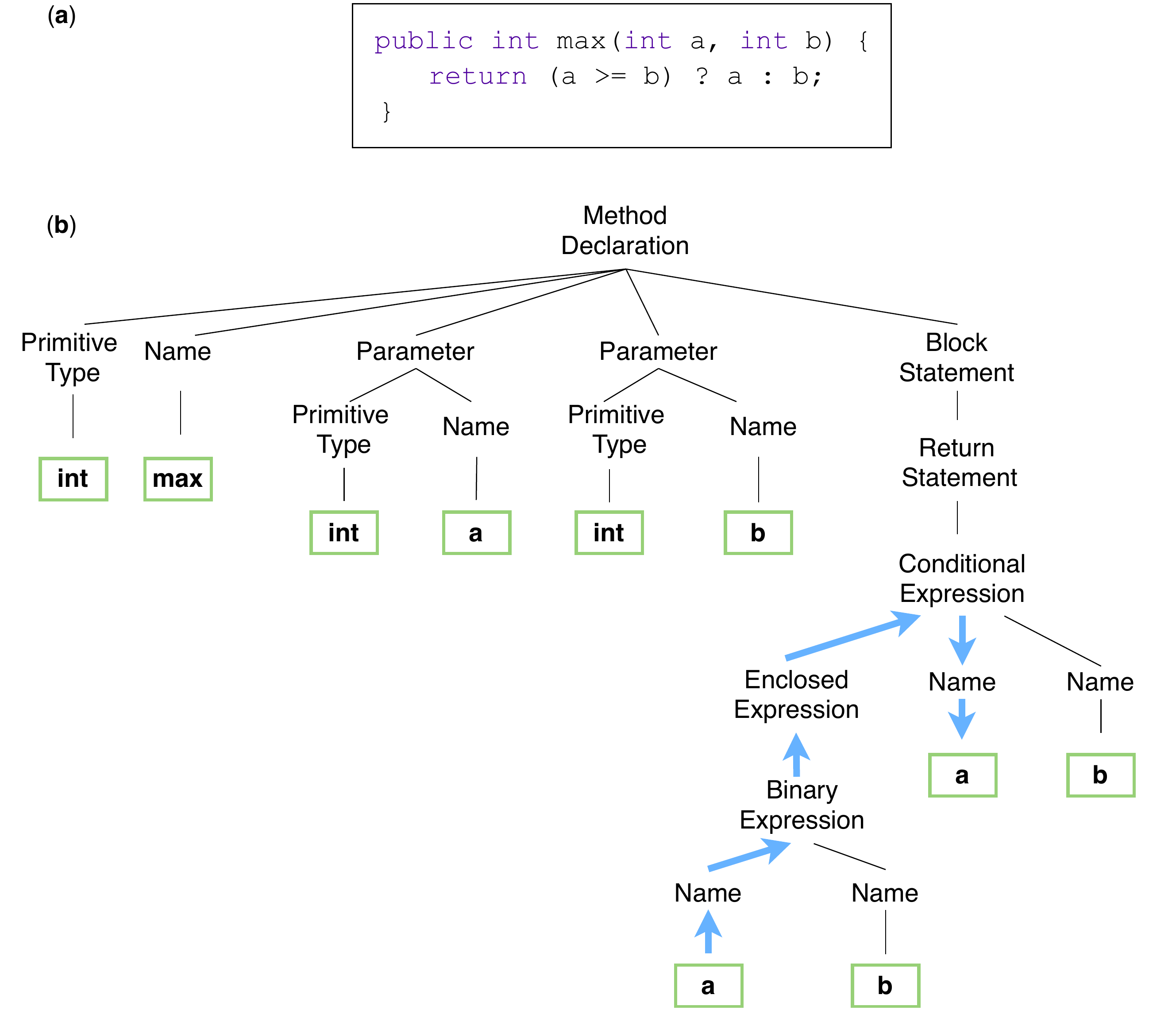}
  \caption{A fragment of code and a path in its AST}
  \label{fig:ast}
  \end{figure}

First of all, the approach parses a snippet of code into an Abstract Syntax Tree (AST) and extracts syntactic paths between all leaf nodes traversing through their lowest common ancestor. Each path is represented as a sequence of intermediate AST nodes between two leaf nodes and arrows which illustrate the direction in the AST. This concept is illustrated in Figure~\ref{fig:ast}: Figure~\ref{fig:ast}(a) shows a code snippet and Figure~\ref{fig:ast}(b) presents a corresponding AST and a path between two nodes in it highlighted in blue. Alon et al. introduce the term $path$-$context$ as a tuple of two nodes and a corresponding path between them. For example, the blue colored path-context from Figure~\ref{fig:ast} is represented as $(a, Name \uparrow BinaryExpression \uparrow EnclosedExpression \uparrow ConditionalExpression \downarrow Name, a)$.

The whole code snippet is considered as a bag of path-contexts. For each component of a path-context (i.e., two leaf nodes and a path between them), the model learns a numeric vector, and then three vectors are concatenated into a single vector called $combined$ $context$ $vector$. To aggregate the information from each path-context into one vector that represents the whole code fragment, the approach employs the attention mechanism that computes a weighted average over all combined context vectors by assigning greater weight to paths that capture more important semantic information for this code fragment. The resulting fixed-length $code$ $vector$ of size 384 can be further used in the prediction task. To reduce the sparsity and the amount of training data, the authors limit the length (i.e., number of nodes) and the width (i.e., maximal allowed difference between sibling nodes) of paths.
 
Based on these embeddings, the authors built a model to predict methods' names from their bodies using a dataset that contains more than 12M methods in Java. As the result, the code2vec model learned to produce embeddings in such a way that similar code snippets are assigned similar embeddings. The authors suggest that these embeddings can be further used for other tasks where this semantics similarity could be useful, such as code summarization, refactoring suggestion, and code search.

In this paper, we investigate the effectiveness of this approach for the refactoring recommendation task: we use the model pre-trained for Java by Alon et al. to build embeddings of classes and movable methods.

\subsection{Representation of data}
\label{subsection:representation}

We apply code2vec to each movable method $m_i$ and each corresponding target class $c_{i_j}$ separately. The code vector for a class $c_{i_j}$ is calculated as an element-wise average of code vectors of methods of this class (except for the method $m_i$ if $c_{i_j}$ is an original class for $m_i$):

\[embedding(c_{i_j})=\dfrac{1}{k} \cdot \sum_{j=1}^{k} embedding(m_j)\]
where $k$ is a number of methods in class $c_{i_j}$.

Then, we concatenate the resulting code vector for the method and the resulting code vector for the class into a single vector of size 768. As a result, we have a set of vectors $v=\{v_{i_1}, v_{i_2}, ..., v_{i_k}\}$, where each of them is represented as:
\[v_{i_j} = [embedding(m_i), embedding(c_{i_j})]\]

To reduce the space dimensionality, we employ Principle Components Analysis (PCA) that transforms a larger number of variables into a smaller number of \textit{principal components}. Reducing the space dimensionality allows us to decrease the operation time of the model training as well as to better manage the available memory by focusing on the most important features. As a result, we got 191 principal components instead of 768. 

\subsection{Generation of the training data}
\label{subsection:training}
One of the main challenges researchers face when implementing machine learning-based approaches is collecting a large quantity of labeled data to train the model. This is especially true for the refactoring recommendations tasks since to our best knowledge there is no publicly available dataset large enough to be used in training deep neural models. In order to solve this problem for the Feature Envy code smell, we used a tool called \textit{MoveMethodGenerator} that was suggested in~\cite{Novozhilov:2019:EMM:3340633.3340640} that automatically generates synthetic datasets suitable for evaluating algorithms that recommend Move Method refactorings. The tool receives a project as input and generates a list of potentially movable methods and target classes where methods can be moved into. The tool filters the list of candidates: for example, static methods, constructors, empty and delegation methods,  methods with no parameters, getters, and setters are not considered as movable methods. The tool is also able to move all potentially movable methods automatically. To automate this process, the tool uses IntelliJ Platform SDK~\footnote{https://www.jetbrains.com/opensource/idea/}. 

To collect training data, we applied MoveMethodGenerator to 1000 open-source Java projects on GitHub that had the most stars. Overall, 18,086 movable methods were detected in 473 projects.  The labeling of the data comes from the assumption that such popular repositories are well-maintained, contain high-quality code and therefore all the methods are placed in correct classes by default. For example, if MoveMethodGenerator concludes that a method $m$ from a class $A$ can be moved to a class $B$ or a class $C$, we add two ``negative'' points to the training data: ``$m$ should not be moved to $B$'' and ``$m$ should not be moved to $C$''. To balance the dataset, we also add two duplicate ``positive'' items: ``method $m$ should be moved to $A$ (original class)''. As a result, the dataset consists of items like \[item = [concat(embedding(m), embedding(C)), label]\] where $concat(a, b)$ is a concatenation of $a$ and $b$, $m$ is a method, $C$ is a target class and $label$ is equal to 1 if the class is a original class for the method, and 0 otherwise.

We split the collected data into three sets: train, test, and validate, in the ratio of 3:1:1 respectively.

\subsection{Classification}
\label{subsection:classifier}
To predict whether a method should be moved to a class or not, we employ an SVM classifier that has demonstrated its effectiveness in code smells detection~\cite{maiga2012smurf, maiga2012support, fontana2017code}. SVM is a technique that uses kernel functions to perform the optimal separation of the data into two categories by a hyperplane. 


To transform SVM's output into probability, we use Platt scaling:
\[Pr(y=1|x)=\dfrac{1}{1+exp(Af(x)+B)}\]
where $Pr \in [0, 1]$ is a probability that $x$ should be moved, $f$ is a uncalibrated output of the SVM, $A$ and $B$ $\in \mathds{R}$ are scalar parameters that are learned by the algorithm.

The approach concludes that a method $m$ should be moved to a class with the highest probability $c_i$ if and only if the probability is greater than the empirically determined value of 0.5. Otherwise, the approach does not recommend a Move Method refactoring for the method $m$ at all, because it is of greater value to suggest fewer refactorings of higher quality (i.e. higher probability). If a class with the highest probability is the original class of the method, we conclude that the method should not be moved anywhere.

A similar technique was described in~\cite{8807230}, our approach to recommending Move Method refactorings is different in the following aspects. Firstly, while their approach only considers embeddings of identifiers obtained using word2vec, our approach considers embeddings of methods' bodies to better capture both their structure and semantic properties. Secondly, our approach does not use any software metrics, whereas their approach employs metrics presented by~\cite{tsantalis2009identification}. We decided to not use metrics to avoid manually defined rules and thresholds.

\section{Evaluation}
\label{section:evaluation}
In this section, we evaluate our approach and compare it to existing refactoring recommendation tools. We decided to evaluate our approach on two datasets: in the first part, we evaluate our approach on the manually-prepared dataset used in~\cite{terra2018jmove}; in the second part, we evaluate our approach on an synthetic dataset. In both cases, we use projects which were not encountered in the training phase and compare our results to the existing recommendation tools, namely JDeodorant~\cite{tsantalis2009identification} and JMove~\cite{terra2018jmove}. They are selected for comparison due to the following reasons: they are publicly available and represent state-of-the-art in this area. JDeodorant is also widely used as a benchmark in prior refactoring recommendation research.

The evaluation consist of the following steps. For each project in our datasets, described in Section~\ref{subsection:datasets}, we search for movable methods and possible target classes using MoveMethodGenerator. Next, we transform the collected data as described in Section~\ref{subsection:representation}. Finally, we run the trained SVM classifier on the preprocessed data from the previous step.

\subsection{Datasets}
\label{subsection:datasets}
Let us now describe the datasets, that we use to evaluate our approach.
\subsubsection{JMove's dataset}

The corpus of Java projects for evaluation was first presented by Terra et al.~\cite{terra2013qualitas}, it is a compiled version of the Qualitas Corpus proposed by Tempero et al.~\cite{tempero2010qualitas}. To evaluate JMove, the authors randomly selected nine projects from the corpus. Then, for each project, the authors manually moved methods from their original classes to randomly selected new classes. Before each move, they made sure that it is possible to move the method back to the original class and that the method has at least four dependencies. The second precondition is explained by the fact that JMove is based on the assumption that methods in well-designed classes usually establish dependencies to similar types and recommends refactorings only for methods that have at least four dependencies (method calls, object instantiations, field accesses, etc).

\begin{table}
    \centering
        \caption{Subject applications from JMove's dataset}
    \label{tab:applications}
    \begin{tabular}{ccccc}
    \toprule
    {\bf Application} & {\bf Version} & {\bf NOC} & {\bf NOM} & {\bf LOC} \\ \midrule
    weka & 3.6.9 &  908 &  16,034 & 257,897\\ 
    ant & 1.8.2 &  760 &  8,586 & 103,402\\ 
    freecol & 0.10.3 & 535 & 6,616 & 93,605\\ 
    jmeter & 2.5.1 &  682 & 7,392 & 81,222\\ 
    freemind & 0.9.0 &  368 &  4,074 & 53,782\\ 
    jtopen & 7.8 &  1,450 &  22,143 & 340,752\\ 
    jhotdraw & 6.2.0 & 520 & 5,878 & 80,153\\ 
    drjava & r5387 &  361 &  4,675 & 88,631\\ 
    maven & 3.0.5 & 154 & 1,568 & 71,065\\
    \bottomrule
\end{tabular}
\end{table}

We evaluate the proposed approach on nine open-source projects from JMove's dataset presented in Table~\ref{tab:applications}. The table contains the information about the project's name, version, number of classes (NOC), number of methods (NOM), and number of lines of source code (LOC).

\subsubsection{Synthetic dataset}
We also evaluate our approach on a synthetic dataset to minimize a possible impact of a manual analysis of samples. We took five high quality open-source projects from a publicly available synthetic dataset called $MoveMethodDataset$, which was compiled using the MoveMethodGenerator tool described in Section~\ref{subsection:training}. The projects used in this dataset are presented in Table~\ref{tab:applications_movemethoddataset} and are different than the projects that were used to train the model (as described in Section~\ref{subsection:training}).

\begin{table}
    \centering
        \caption{Subject applications from the synthetic dataset}
    \label{tab:applications_movemethoddataset}
    \begin{tabular}{ccccc}
    \toprule
    {\bf Application} & {\bf Version} & {\bf NOC} & {\bf NOM} & {\bf LOC} \\ \midrule
    pmd & 6.13.0 & 1,147 & 8,637 & 119,430\\
    cayenne & 4.2 & 1,499 & 12,164 & 275,450\\ 
    pinpoint & 1.9.0 & 2,551 & 17,024 & 290,974\\ 
    jenkins & 1.51 & 768 & 6,292 & 155,667\\
    drools & 7.22.0 & 2.758 & 27,793 & 680,234\\
    \bottomrule
\end{tabular}
\end{table}

\subsection{Evaluation Metrics}
In this work, we use three widely used statistical metrics to evaluate the effectiveness of approaches, namely, precision, recall and $F_1$ score. 

We calculate precision as a ratio between the number of the correct recommendations and the number of all recommendations provided by the tool, recall as a ratio between the number of correct recommendations and the number of moved methods in a project, and $F_1$ score is calculated as a harmonic mean of the precision and recall values.

\begin{align*}
\text{Precision}  &= \frac{\# \: \text{of correct refactorings}}{\# \: \text{of  recommended  refactorings}} \\[1ex]
\text{Recall} &= \frac{\# \: \text{of correct refactorings}}{\# \: \text{of  moved  methods}} \\[1ex]
F_1  &= 2 \times \frac {\text{Precision} \times \text{Recall}}{\text{Precision}+\text{Recall}}
\end{align*}

The recommended Move Method refactoring for a method is considered correct if it suggests to move the method back to its original class because we assume that projects used in evaluation also contain a high-quality code.

\begin{table*}
    \centering
        \caption{Evaluation results on JMove's dataset}
    \label{tab:evaluation_results_jmove}
  \begin{tabular}{l|lll|lll|llll}
    \toprule
    \multirow{2}{*}{Application} &
      \multicolumn{3}{c|}{Proposed approach} &
      \multicolumn{3}{c|}{JDeodorant} &
      \multicolumn{3}{c}{JMove} \\
      & \textit{Precision} & \textit{Recall} & \textit{\: \: $F_{1}$} & \textit{Precision} & \textit{Recall} & \textit{\: \: $F_{1}$} &\textit{    Precision} & \textit{Recall} & \textit{\: \: $F_{1}$}\\
      \midrule
    weka & \bf 0.224 & 0.645 & \bf 0.332 & 0.059 & 0.548 & 0.107 & 0.108 & \bf 0.741 & 0.189 \\ 
    ant &  \bf 0.197 & 0.56 & \bf 0.292 & 0.171 & 0.48 & 0.252 & 0.173 & \bf 0.84 & 0.287\\ 
    freecol &  0.048 & 0.647 & 0.089 & 0.030 &0.294 & 0.054 & \bf 0.074 & \bf 0.764 & \bf 0.135\\ 
    jmeter  &  0.264 & 0.36 & 0.305 & 0.236 & 0.52 & 0.325 & \bf 0.275 & \bf 0.76 & \bf 0.404\\
    freemind  &  \bf 0.8 & 0.333 & \bf 0.470 & 0.166 & 0.583 & 0.258 & 0.148 & \bf 0.666 & 0.242\\
    jtopen  &  \bf 0.416 & 0.512 & \bf 0.459 & 0.207 & 0.447 &  0.283 & 0.208 & \bf 0.894 & 0.337\\ 
    jhotdraw & \bf 0.470  & 0.4 & 0.432 & 0.454 & 0.5 & 0.476 & 0.468 & \bf 0.75 & \bf 0.576\\ 
    drjava &  \bf 0.428 & 0.5 & \bf 0.461 & 0.128 & 0.555 & 0.208 & 0.128 & \bf 0.777 & 0.22\\ 
    maven & \bf 0.545 & \bf 0.541 & \bf 0.543 & 0.139 &0.25 &  0.179 & 0.104 & 0.375 & 0.163\\ 
    \bf Average &  \bf 0.376 & 0.5 & \bf 0.375 & 0.177 & 0.464 & 0.238 & 0.187 & \bf 0.730 & 0.284\\
    \bottomrule
  \end{tabular}
\end{table*}

\begin{table*}
    \centering
        \caption{Evaluation results on MoveMethodDataset}
    \label{tab:evaluation_results_movemethoddataset}
  \begin{tabular}{l|lll|lll}
    \toprule
    \multirow{2}{*}{Application} &
      \multicolumn{3}{c|}{Proposed approach} &
      \multicolumn{3}{c}{JMove} \\
      & \textit{Precision} & \textit{Recall} & \textit{\: \: $F_{1}$} & \textit{Precision} & \textit{Recall} & \textit{\: \: $F_{1}$}\\
      \midrule
    pmd & \bf 0.238 & \bf 0.446 & \bf 0.310 & 0.076 & 0.021 & 0.033 \\
    cayenne & \bf 0.437 & \bf 0.181 & \textbf{0.256} & 0.029 & 0.060 & 0.039\\
    pinpoint & \bf 0.162 & \bf 0.214 & \bf 0.184 & 0.115 & 0.143 & 0.127\\
    jenkins & \bf 0.259 & \bf 0.538 & \bf 0.35 & 0.028 & 0.307 & 0.051\\
    drools & \bf 0.242 & \bf 0.384 & \bf 0.297 & 0.121 & 0.173 & 0.142\\
    \bf Average & \bf 0.268 & \bf 0.353 & \bf 0.28 & 0.074 & 0.140 & 0.138\\
    \bottomrule
  \end{tabular}
\end{table*}

\subsection{Results}
The results of the evaluation on JMove's dataset are presented in Table~\ref{tab:evaluation_results_jmove}. The results show that our approach demonstrates the highest average precision and $F_1$ score of 0.376 and 0.375 respectively among all approaches whereas JMove demonstrates the highest recall of 0.730. Thus, our approach achieves an increase of 32\% in $F_1$ compared to JMove, and an increase of 57\% compared to JDeodorant.
In our experience, refactoring recommendation is a task where precision is a more relevant indicator of the approach's effectiveness for real-world use: it is more important to suggest correct and reasonable refactorings rather than suggest \textit{more} refactorings of lower quality that developers will waste precious time browsing through when using the tool.

We believe that the reason that our approach shows better precision has to do with employing the path-based representation of code. It would seem that taking into account the semantic qualities of code and semantic similarity between methods and classes leads to better recommendations than using various software metrics.

Evaluation results on the synthetic dataset are presented in Table~\ref{tab:evaluation_results_movemethoddataset}. It can be seen that the proposed approach outperforms JMove: the average precision and $F_1$ score of JMove are 0.074 and 0.138 respectively, whereas our approach demonstrates the average precision of 0.268 (3.6-fold increase) and $F_1$ score of 0.28 (2-fold increase). JDeodorant did not suggest correct recommendations in this case due to the fact that it recommends moving a method from the original class to another only if the method accesses more members of another class than the original one. MoveMethodGenerator generates a dataset by moving a candidate method from its original class to a class of one of the arguments of this method. Thus, the method also accesses the members of a new class. Moreover, JDeodorant recommends refactorings only if the refactoring improves a specific metric based on coupling and cohesion, which might not be true in case of the automatically generated dataset.

The synthetic nature of the dataset did not impact the value of structural and semantic information of the code that is captured by path-based embeddings, and therefore our approach still shows better results than heuristics-based techniques.

\section{Threats to Validity}
\label{section:threats}

Our work deals with several threats to validity.

From the standpoint of the proposed pipeline, our approach is based on the assumption that the code in the popular projects is high quality, meaning that all methods are located correctly, which might not always be the case. Based on this assumption, we build our training data as a balanced mixture of corresponding positive and negative examples. If the assumption is wrong, this might lead to a bias in the model training. However, as one of the recent studies show, the proportion of smelly code in software systems is usually rather small~\cite{palomba2018diffuseness}. Besides, the approach of using the most popular projects on GitHub for training is very often used this area of studies.

In this study, we only use SVM and do not experiment with other classifiers. However, prior studies have shown that in the refactoring recommendation task, the choice of the classifier is not as significant as the selection of features~\cite{fontana2016comparing, di2018detecting}.

From the standpoint of the evaluation, we limit ourselves to a relatively small sample of projects. However, this sample contains projects that come from different domains and our approach demonstrates consistent results on all of them. A more thorough evaluation is a subject for future work.

Overall, while these threats to validity are important to note, we believe that they do not invalidate the proposed approach, but point the way to the possible directions of improving the described technique and refactoring recommendation field in general.

\section{Conclusion and future work}
\label{section:conclusion}
In this paper, we propose an approach to recommend Move Method refactorings that relies on the path-based representation of code that is used to train a machine learning classifier. Such representation allows us to capture semantics of the code in such a way that similar pieces of code are mapped to similar vectors, and having such embeddings allows us to train a classifier. For each movable method, the approach suggests to either move the method to a more semantically similar class or leave it in its original class.

We evaluate the approach on two datasets: the first one consists of several open-source projects with manually moved methods and the second one consists of projects with automatically injected Feature Envy smell. On the hand-crafted dataset, the average $F_1$ score of JDeodorant is 0.238, JMove is 0.284, whereas our approach demonstrates the average $F_1$ score of 0.375. On the synthetic dataset, the average $F_1$ score of JMove is 0.138, whereas our approach demonstrates the $F_1$ score of 0.28. The evaluation results show that the proposed approach is able to recommend appropriate refactorings in both cases and outperforms state-of-the-art tools. Path-based representation seems to play a key role in the effectiveness of the method.

The research can be extended in several directions, here are some of them:
\begin{itemize}
    \item The proposed approach can be further developed and perfected by experimenting with different classifiers, gathering larger amount or different kind of training data, and evaluating on different datasets.
    
    \item It might be of interest to investigate the effectiveness of the proposed approach in application to other code smells. For example, detection of Misplaced Class code smell might be also resolved as a classification task. This smell relates to the situation when the class is placed in a package with classes that it is not related to and can be solved by moving the class to a more related package.
    
    \item The method can also be extended to other programming languages. Currently, the approach works only with code written in Java for two reasons: (1) MoveMethodGenerator searches for possible movable methods in programs written in Java, (2) we use the code2vec model that is pre-trained on the Java corpus by its authors. The possible way to extend the proposed approach to other languages is the generalization of MoveMethodGenerator and retraining code2vec (which seems to be a fairly simple technical task using a tool for mining path-based representations in different languages proposed by Kovalenko et al.~\cite{kovalenko2019pathminer}).
    
    \item Another possibility is the development of an automatic Move Method refactoring recommendation tool as a plug-in for an IDE. For instance, the IntelliJ Platform provides an infrastructure for developing plug-ins to popular IDEs such as IntelliJ IDEA, PyCharm, CLion, etc. For example, the proposed approach might be implemented as a plug-in for IntelliJ IDEA, a popular IDE for Java developers.
\end{itemize}

Overall, code smells continue to be an important issue in the modern software development and require further research. One might expect that the combination of semantically-aware representations of code and various machine learning techniques, similar to the one proposed in this work, can be a promising approach to detect code smells in code and suggest appropriate refactorings to resolve them.

\bibliographystyle{plain}
\bibliography{bibliography.bib}
\end{document}